\documentclass{article}
\usepackage{spconf,amsmath,graphicx,hyperref, amssymb, amsfonts}
\usepackage{tabularray}
\usepackage{xcolor}
\usepackage{bbm}
\usepackage{svg}

\ninept

\newcommand{\myparagraph}[1]{\noindent\textbf{#1} \hspace{.12cm}}

\title{Modeling strategies for speech enhancement in the latent space of a neural audio codec}

\name{Sofiene Kammoun$^{\star}$ \qquad Xavier Alameda-Pineda$^{\dagger}$ \qquad Simon Leglaive$^{\star}$}
  \address{$^{\star}$CentraleSupélec, IETR (UMR CNRS 6164), France \\
      $^{\dagger}$Inria at Univ. Grenoble Alpes, CNRS, LJK, France} 

\begin{document}
\maketitle
\begin{abstract}
    Neural audio codecs (NACs) provide compact latent speech representations in the form of sequences of continuous vectors or discrete tokens. In this work, we investigate how these two types of speech representations compare when used as training targets for supervised speech enhancement. We consider both autoregressive and non-autoregressive speech enhancement models based on the Conformer architecture, as well as a simple baseline where the NAC encoder is simply fine-tuned for speech enhancement. Our experiments reveal three key findings: predicting continuous latent representations consistently outperforms discrete token prediction; autoregressive models achieve higher quality but at the expense of intelligibility and efficiency, making non-autoregressive models more attractive in practice; and adding encoder fine-tuning yields the strongest enhancement metrics overall, though at the cost of degraded codec reconstruction. The code and audio samples are available online.\footnote{\texttt{sofienekammoun.github.io/SE-NAC-25/}}
\end{abstract}

\section{Introduction}

Speech enhancement (SE) addresses the problem of estimating a clean speech signal from a degraded recording affected by additive noise, reverberation, or other distortions. Although the target output is often the clean waveform (for listening or downstream tasks), the enhancement process can be defined in different representation domains. The choice of representation (waveform, time–frequency, continuous latent vectors, or discrete tokens) strongly conditions model design, inference speed, and the quality and intelligibility of the restored signal.

Classic SE algorithms worked predominantly in the short-time Fourier transform (STFT) domain, exploiting structure in the time–frequency plane and statistical assumptions about speech and noise \cite{boll1979suppression}. Early deep learning approaches to SE followed the established time–frequency pipeline by estimating time-frequency masks or magnitudes using discriminative training of neural networks \cite{ narayanan2013ideal, weninger2015speech}. Later, time-domain architectures such as Conv-TasNet \cite{luo2019conv} showed that learned encoder–separator–decoder pipelines with 1D temporal convolutions can outperform oracle time-frequency masking in speech separation, which removes the need for explicit phase estimation by using trainable filterbanks and time-domain loss functions. The effectiveness of this time-domain approach was also confirmed for SE \cite{defossez2020real}.

In parallel, neural audio codecs (NACs), initially introduced in \cite{zeghidour2021soundstream}, are being adopted across speech processing frameworks, where they provide an alternative representation space for speech signals~\cite{Investigating_NAC}. NACs compress audio through an encoder-decoder framework into compact latent sequences represented as continuous vectors or as discrete tokens produced by residual vector quantization (RVQ) \cite{juang1982multiple}. The appeal of these representations for downstream tasks is due to two key advantages. First, they offer a very compressed representation of the data, potentially reducing memory and compute for sequence models. Second, the discrete-token view makes it straightforward to reuse modeling techniques from natural language processing (NLP).

Recent audio generative models such as \cite{VALLE} process audio signals as sequences of discrete tokens, using RVQ-aware and often autoregressive (AR) transformer-based architectures \cite{atten2017}. Inspired by this success, recent works have developed SE methods that generate clean speech tokens conditioned on noisy speech tokens \cite{Genhancer,masksr,gense}. While transformers are often assumed to require discrete tokens as input, their first layer is an embedding layer that produces continuous representations, making it feasible to directly use the continuous latent representation of the NAC without modifying the rest of the model. Recent works have started exploring the use of transformer-based architectures using continuous representations for image generation \cite{li2024autoregressive}, speech synthesis \cite{meng-etal-2025-autoregressive}, speech separation \cite{Sep_cont2024}, and SE \cite{SE_cont2025,sun2025efficient}. 

Moreover, in general audio generation tasks, AR models are appropriate when the output length or alignment between input and output is unknown, which is the case for tasks like text-to-speech (TTS) \cite{VALLE}, prompt-based music \cite{copet2023simple}, and image \cite{RQ-trans}  generation. However, in SE, the input and output signals (noisy and clean speech) are of equal length and perfectly aligned, with no absolute need for diverse predictions. This reduces the intrinsic need for AR modeling. Still, AR models offer the advantage of modeling intra-sequence temporal dependencies. In contrast, non-AR models generate the entire sequence simultaneously, relying only on input-output dependencies. Recent concurrent studies have also begun to explore this trade-off for SE \cite{della2025autoregressive, yang2025investigating}, offering complementary insights. 

Finally, SE and codecs are two essential components of voice communication systems, making the study of their interaction practically relevant. A natural approach to SE using NACs is to fine-tune the NAC encoder to produce a clean speech latent representation directly from the noisy input. However, this may compromise the NAC's reconstruction performance. To our knowledge, this simple strategy has only received very little attention in prior work~\cite{hauret2025real}.

In this work, we focus on three central questions regarding SE using NACs: (i) whether to model discrete tokens or continuous vectors in the NAC's latent space, (ii) how AR and non-AR models compare, and (iii) how the fine-tuning of the NAC's encoder compares to adding a separate SE model. Our goal is not to optimize any single model for peak performance but to provide a fair comparative study of these signal representation spaces and modeling choices.

\vspace{0pt}
\section{Method}
\vspace{0pt}

We frame our study of SE in NAC latent spaces along three modeling axes introduced above. To structure the discussion, we design a family of models that systematically vary along these axes while keeping other components comparable. 
All methods share the following common structure. A pretrained NAC maps noisy and clean waveforms into latent representations, either continuous embeddings or discrete tokens. Enhancement models then learn a conditional distribution of clean latents given noisy latents. Depending on the variant, the predictor may operate autoregressively over time or non-autoregressively, targeting either continuous vectors or discrete tokens. Finally, the NAC decoder reconstructs the waveform from the predicted latent sequence. We also compare these models to a simple fine-tuning of the NAC encoder. In the following, we denote by $\theta$ the model parameters for each SE method. All models are trained in a supervised manner by minimizing the negative log-likelihood. 
\vspace{0pt}
\subsection{Neural audio codecs}

NACs are encoder-decoder architectures built with 1D convolutional networks, but they further introduce RVQ of the latent representation extracted by the encoder \cite{zeghidour2021soundstream}. Let $\mathbf{y}_{\text{wav}}$ and $\mathbf{x}_{\text{wav}} \in \mathbb{R}^{d \cdot f_s}$ denote the noisy and clean speech waveforms of duration $d$ in seconds and sampling rate $f_s$ in Hertz, assuming $d \cdot f_s \in \mathbb{N}$ without loss of generality. A NAC can be described with the following operations:
\begin{equation}
\label{NAC}
  \begin{aligned} 
 \text{Tokenize: }&
  \begin{aligned}
\mathbf{x}_{\text{wav}} \\
\mathbf{y}_{\text{wav}}
\end{aligned}
 \overset{\displaystyle\mathcal{E}}{\longrightarrow} 
\begin{aligned}
\bar{\mathbf{x}} \\
\bar{\mathbf{y}}
\end{aligned} 
  \in \mathbb{R}^{L \times T}  \overset{\displaystyle\mathcal{Q}_{\mathcal{C}}}{\longrightarrow} 
\begin{aligned}
\mathbf{x} \\
\mathbf{y}
\end{aligned} 
\in \{1,...,K\}^{N \times T}, \\
 \text{Detokenize: }&
\begin{aligned}
\mathbf{x} \\
\mathbf{y}
\end{aligned} 
  \overset{\displaystyle\mathcal{Q}_{\mathcal{C}}^{-1}}{\longrightarrow} 
\begin{aligned}
\hat{\bar{\mathbf{x}}}  \\
\hat{\bar{\mathbf{y}}} 
\end{aligned} 
\in \mathbb{R}^{L \times T} \overset{\displaystyle\mathcal{D}}{\longrightarrow} 
\begin{aligned}
\hat{\mathbf{x}}_{\text{wav}} \\
\hat{\mathbf{y}}_{\text{wav}} 
\end{aligned} 
\in \mathbb{R}^{d \cdot f_s},
\end{aligned} 
\end{equation}
where $\mathcal{E}$ is the encoder that encodes the waveform into an $L$-dimensional downsampled continuous representation of length $T$, $\mathcal{D}$ is the decoder that performs the reverse operation, and $\mathcal{Q}_{\mathcal{C}}$ and $\mathcal{Q}_{\mathcal{C}}^{-1}$ denote the direct and inverse RVQ modules, which depend on a set of codebooks $\mathcal{C} = \{\mathcal{C}_n \in \mathbb{R}^{L \times K} \}_{n=1}^N$, with $K$ codebook vectors of dimension $L$ for each quantization stage $n \in \{1,..., N\}$. See \cite{zeghidour2021soundstream} for a description of RVQ \cite{juang1982multiple} in the context of NACs.

\vspace{0pt}
\subsection{Speech enhancement models}

\subsubsection{Discrete token representation} 

We first consider modeling strategies where the noisy and clean speech signals are represented as discrete tokens extracted using a pretrained NAC: $\mathbf{x} = \{x_{t,n}\}_{t=1,n=1}^{T, N} \in \{1,..., K\}^{N\times  T}$, $\mathbf{y} = \{y_{t,n}\}_{t=1,n=1}^{T, N} \in \{1,..., K\}^{N\times  T}$.\\

\myparagraph{Discrete autoregressive model (D-AR)} This first SE model is AR over time and quantization depth, and can be written using the chain rule of probabilities as:  
\begin{equation}
\begin{aligned}
    p_\theta&(\mathbf{x} \mid \mathbf{y}) = \prod_{t=1}^T \prod_{n=1}^N p_\theta(x_{t,n} \mid \mathbf{x}_{1:t-1}, \mathbf{x}_{t, 1:n-1}, \mathbf{y}) \\ &= \prod_{t=1}^T \prod_{n=1}^N \prod_{k=1}^K f^{\texttt{(D-AR)}}_{t,n,k}(\mathbf{x}_{1:t-1} , \mathbf{x}_{t, 1:n-1}, \mathbf{y} ; \theta)^{\mathbbm{1}\{x_{t,n} = k\}},
\end{aligned}
\label{D-AR}
\end{equation}
where $\mathbbm{1}\{\cdot\}$ denotes the indicator function, and $f^{\texttt{(D-AR)}}_{t,n,k}(\cdot) \in [0,1]$ denotes the probability that the discrete clean speech token $x_{t,n}$ at time $t$ and quantization stage $n$ equals $k \in \{1,...,K\}$, given the noisy tokens $\mathbf{y}$, the clean speech tokens $\mathbf{x}_{t, 1:n-1}$ at time $t$ and quantization stages $1$ to $n-1$, and the past clean speech tokens $\mathbf{x}_{1:t-1}$ at all quantization stages. Therefore, we have $\sum_{k=1}^K f_{t,n,k}(\cdot) = 1$. Inspired by the RQ-Transformer model \cite{RQ-trans} for text-to-image generation and by the Conformer model \cite{conformer} for speech recognition, in this work, we propose the RQ-Conformer to implement the function $f^{\texttt{(D-AR)}}$. The proposed model is composed of $3$ modules: (i) a bidirectional Conformer for processing the noisy speech tokens $\mathbf{y}$, (ii) a causal Conformer to autoregressively process the clean speech tokens over time $\mathbf{x}_{1:t-1}$, and (iii) a causal transformer processing the outputs of both previous modules and the clean speech tokens over the quantization-depth dimension $\mathbf{x}_{t, 1:n-1}$ providing the predicted output $f^{\texttt{(D-AR)}}_{t,n,k}$. This choice is motivated by the recurrent nature of RVQ, where the quantization at level $n$ depends on the quantization at level $n-1$. At the input, a trainable embedding layer embeds each discrete token in $\{1,..., K \}$ into a continuous vector of dimension $H$ (the internal Conformer dimension).

\vspace{4pt}
\myparagraph{Discrete non-autoregressive model (D-NAR)} We now remove the AR dependencies in the previous model described in \eqref{D-AR}, such that:
\begin{equation}
\label{D_NAR}
p_\theta(\mathbf{x} \mid \mathbf{y}) = \prod_{t=1}^T \prod_{n=1}^N \prod_{k=1}^K f^{\texttt{(D-NAR)}}_{t,n,k}(\mathbf{y} ; \theta)^{\mathbbm{1}\{x_{t,n} = k\}},
\end{equation} 
where $f^{\texttt{(D-NAR)}}: \{1,..., K\}^{N \times T} \mapsto [0,1]^{T \times N \times K}$ is now simply composed of a bidirectional Conformer model that processes the noisy tokens $\mathbf{y}$ in a feed-forward manner. At the output, the model contains $N$ classification heads that convert the Conformer's output to a sequence of probabilities over $\{1,..., K\}$. \vspace{-8pt}

\subsubsection{Continuous latent representation}  
\label{Cont}

We now consider modeling strategies where the noisy and clean speech signals are represented as continuous latent vectors extracted from a pretrained NAC, i.e., $ \bar{\mathbf{x}} = \{\bar{\mathbf{x}}_t\}_{t=1}^T, \bar{\mathbf{y}}  = \{\bar{\mathbf{y}}_t\}_{t=1}^T \in \mathbb{R}^{L \times T}$. So, at a time step $t$, instead of predicting a categorical distribution over the discrete tokens for each quantization stage $n$, the model will directly provide a prediction of the continuous vector $\bar{\mathbf{x}}_t$, given $\bar{\mathbf{y}}$ and potentially additional conditions. Additionally, the input embedding layer becomes a simple linear layer that projects input vectors from dimension $L$ to $ H$.

\vspace{4pt}
\myparagraph{Continuous autoregressive model (C-AR)} The first model in the continuous case is autoregressive and can be written as:
\begin{equation}
    \begin{aligned}
    p_\theta(\bar{\mathbf{x}} \mid \bar{\mathbf{y}}) &= \prod_{t=1}^T p_\theta(\bar{\mathbf{x}}_t \mid  \bar{\mathbf{x}}_{1:t-1}, \bar{\mathbf{y}})\\ &= \prod_{t=1}^T \mathcal{N}\left(\bar{\mathbf{x}}_t ; f^{\texttt{(C-AR)}}_t( \bar{\mathbf{x}}_{1:t-1}, \bar{\mathbf{y}} ; \theta), \mathbf{I}_L \right),
    \end{aligned}
    \label{C-AR}
\end{equation}
where $\mathcal{N}$ denotes a multivariate Gaussian distribution over $\bar{\mathbf{x}}_t$ with mean vector $f^{\texttt{(C-AR)}}_t(\cdot)$ and identity covariance matrix $\mathbf{I}_L$. Here, $f$ denotes a causal Conformer model for processing the noisy speech embeddings $\bar{\mathbf{y}}$ concatenated to the clean speech latent vectors $\bar{\mathbf{x}}_{1:t-1}$. In this approach, we consider temporal dependencies using an AR model over time. However, we assume that all dimensions of the continuous latent vectors $\bar{\mathbf{x}}_t$ are independent.

\vspace{3pt}
\myparagraph{Continuous non-autoregressive model (C-NAR)} As in the discrete case, we now remove the autoregressive dependencies in the previous model described in \eqref{C-AR}, such that:
\begin{equation}
p_\theta(\bar{\mathbf{x}} \mid \bar{\mathbf{y}}) = \prod_{t=1}^T p_\theta(\bar{\mathbf{x}}_t \mid \bar{\mathbf{y}}) = \prod_{t=1}^T \mathcal{N}\left(\bar{\mathbf{x}}_t ; f_t^{\texttt{(C-NAR)}}(\bar{\mathbf{y}} ; \theta), \mathbf{I}_L \right),
\label{C-NAR}
\end{equation}
where $f^{\texttt{(C-NAR)}}:  \mathbb{R}^{L \times T} \mapsto \mathbb{R}^{L \times T}$ denotes a bidirectional Conformer model, with the same input layer previously described and a predictive head that generates the mean vector of the Gaussian distribution. 
% \vspace{-6pt}
\subsubsection{Baseline: fine-tuning the encoder}

In contrast to the approaches described above, we also consider a simpler baseline where the NAC encoder is fed with the noisy speech waveform and fine-tuned to provide the clean speech latent representation, bypassing the need for an additional sequence model. 
In \textit{the continuous representation} setting (\textbf{C-FT} model), the encoder is fine-tuned to predict the clean speech continuous latent representation. The formulation of this model closely resembles \eqref{C-NAR} with $\bar{\mathbf{y}}$ replaced by $\mathbf{y}_{\text{wav}}$. In \textit{the discrete representation} setting (\textbf{D-FT} model), the target is the sequence of clean speech discrete tokens. To bridge the gap between continuous encoder outputs and discrete tokens, we follow the so-called ``soft labeling" strategy described in \cite{RQ-trans}. Specifically, the encoder output vectors are mapped to probability distributions over codebook indices by measuring their Euclidean distance to each codebook vector and normalizing using a softmax function. We use the straight-through estimator for gradient backpropagation \cite{VQVAE}.
% \vspace{-7pt}
\subsection{Training and inference}

All models considered here are trained in a supervised fashion on paired noisy–clean speech data by maximizing the conditional likelihood of clean speech given noisy speech, which is simply defined by taking the logarithm of equations \eqref{D-AR} to \eqref{C-NAR}. When the model is defined over the continuous representation $\bar{\mathbf{x}}$, maximizing the likelihood is equivalent to minimizing the mean squared error (MSE). In the other case, when the model is defined over the discrete latent representation $\mathbf{x}$, this amounts to minimizing the cross-entropy loss. For both AR models, teacher forcing is used during training. For the \textbf{C-AR} model, the input vectors $\bar{\mathbf{x}}_{1:t-1}$ are quantized using the NAC's RVQ module before feeding them to the Conformer model, in order to mitigate the problem of error accumulation that usually occurs for AR models.

Independent of the representation space, given a trained model, SE is achieved by taking the argmax over $\mathbf{x}$ or $\bar{\mathbf{x}}$ of the above-defined probabilistic models. In this work, we are not interested in introducing stochasticity in the SE process by sampling the distributions. We would rather estimate the most likely clean speech signal for a given trained model and noisy signal. For continuous-prediction models, the estimated vectors are quantized before being passed to the NAC decoder, ensuring compatibility with the codec reconstruction pipeline, and for discrete-prediction models, the predictions go through the detokenization process described in \eqref{NAC}. For simplicity, we segment noisy speech audio into chunks of $1$ second and process them separately for all of our experiments.
% \vspace{-8pt}
\section{Experiments}

\begin{table*}[ht]
\centering
\setlength{\tabcolsep}{2pt}
\renewcommand{\arraystretch}{1} 
\fontsize{9pt}{10pt}\selectfont 
\scalebox{0.9}{%
\begin{tabular}{cccccccccc}
Model & OVRL ($\uparrow$) & SIG ($\uparrow$) & BAK ($\uparrow$) & P808 ($\uparrow$) & UTMOS ($\uparrow$) & CosSim ($\uparrow$) & dWER ($\downarrow$) & GFLOPs ($\downarrow$) & Parameters (M) \\[1pt]
\hline\\[-7pt]
DCCRNet \cite{dccrn} & 2.80 & 3.15 & 4.03 & 3.54 & 3.01 & 96.6 & 11.80 & 26 & 3.7 \\
DCUNet \cite{dcunet} & 2.97 & 3.33 & 3.96 & 3.62 & 3.07 & 96.7 & \textbf{10.11} & 250 & 7.7 \\
ConvTasNet \cite{luo2019conv} & 3.11 & 3.39 & 4.01 & 3.31 & 3.27 & 96.4 & 11.30 & 10 & 5.0\\
DPTNet \cite{dptnet} & 3.00 & 3.32 & \underline{4.05} & 3.31 & 3.38 & 96.6 & 10.75 & 2 & 2.8 \\
AnCoGen \cite{sadok2025ancogen} & 3.00 & 3.32 & \underline{4.05} & 3.31 & 3.38 & 96.6 & 19.30 & - & -  \\[1pt]
\hline\\[-7pt]
D-AR & 2.90 & 3.17 & 3.99 & 3.53 & 2.76 & 95.2 & 25.09 & 5857 & 82.2 \\
D-NAR & 2.89 & 3.18 & 3.94 & 3.50 & 2.72 & 95.5 & 23.12 & 6 & 68.7\\
D-NAR$\star$ & 2.91 & 3.20 & 3.94 & 3.51 & 2.80 & 96.4 & 15.93 & 6 & 64.3  \\[1pt]
\hline\\[-7pt]
C-AR & \textbf{3.32} & \textbf{3.61} & \textbf{4.07} & \textbf{3.77} & \textbf{3.61} & 96.2 & 20.47 & 472 & 63.6\\
C-NAR & \underline{3.25} & \underline{3.56} & 4.01 & 3.60 & 3.54 & \underline{97.0} & 13.48 & 6 & 62.5 \\[1pt]
C-NAR-FT & 3.24 & \underline{3.56} & 4.03 & \underline{3.67} & \underline{3.60} & \textbf{97.2} & \underline{11.07} & 6 & 62.5 + 21.5 \\[1pt]
\hline\\[-7pt]
D-FT & 2.84 & 3.12 & 3.97 & 3.40 & 2.63 & 95.3 & 24.42 & 0 & 21.5 \\
C-FT & 3.20 & 3.52 & 4.00 & 3.58 & 3.37 & 96.8 & 12.81 & 0 & 21.5 \\[1pt]
\hline\\[-7pt]
STFT-NAR & 2.69 & 3.10 & 3.63 & 3.06 & 2.54 & 95.0 & 20.69 & 6 & 64.8  \\[1pt]
\hline\\[-7pt]
Noisy speech & 1.75 & 2.46 & 1.81 & 2.62 & 1.51 & 93.9 & 30.00 & \\
\hline
\end{tabular}
}
\caption{SE results (best and second-best scores in each column are bold and underlined).}
\vspace{-6pt}
\label{RES1}
\end{table*}

\subsection{Experimental setup}

\myparagraph{Dataset} Libri1Mix \cite{librimix} is a single-speaker noisy speech dataset constructed from clean utterances of LibriSpeech \cite{librispeech} combined with noise samples from WHAM! \cite{wham}. It covers a range of signal-to-noise ratios (SNRs) from $-6$ to $3$ dB, simulating challenging noisy conditions. We use the $\texttt{train-360}$ subset for training, which contains $156$ hours of paired noisy/clean speech, and the test and dev subsets for validation and testing, which each contain $4$ hours.

\myparagraph{NAC} For the NAC model, we resort to Descript Audio Codec (DAC) \cite{DAC}, widely used across similar lines of work. We use the 16~kHz variant, which has $N=12$ quantization stages, with each codebook containing $K=1024$ vectors of dimension size $L=1024$, and $T=50$ for one second of audio. This model is trained on an extensive dataset of speech, music, and environmental sounds.

\myparagraph{Hyperparameters}
All enhancement models are based on the Conformer architecture \cite{conformer} with hidden dimension $H=384$. The hyperparameters are chosen such that all models contain approximately $6$–$8 \times 10^7$ parameters, on the same order as the NAC encoder. For the continuous models, we use a 16-layer Conformer with a linear prediction head projecting from $H$ to the codec's continuous representation dimension $L$. For the discrete models, the D-AR variant consists of an 8-layer bidirectional Conformer processing noisy tokens, an 8-layer causal Conformer autoregressively modeling past clean tokens, and a 6-layer causal Transformer along the quantization-depth dimension. The D-NAR model uses an 8-layer Conformer followed by $N=12$ feed-forward prediction heads, each projecting to a distribution over the corresponding codebook indices. Training is performed on one-second paired speech segments, with a batch size of $32$ per GPU (4$\times$ NVIDIA HGX A100)  for $300$ epochs. We use AdamW with $\beta=(0.9, 0.95)$, weight decay $0.05$, and a cosine learning rate schedule with $10$ warm-up epochs. Following common scaling strategies, the maximum learning rate is set to $0.005 \times (\text{batch size}/256)$.

\myparagraph{Metrics}
To evaluate the models' performance, we rely on non-intrusive quality metrics commonly used in the literature on generative SE. For speech quality, we use the DNSMOS P.$835$ \cite{reddy2022dnsmos}, a non-intrusive predictor trained to approximate human ratings. DNSMOS provides scores along three dimensions: SIG (speech quality), BAK (background noise suppression), and OVRL (overall quality). In addition, we include DNSMOS P.$808$ \cite{reddy2021dnsmos} as a complementary quality estimator. To assess naturalness, we use UTMOS \cite{saeki2022utmos}, a neural MOS predictor particularly sensitive to artifacts introduced by vocoding or compression, and focuses more on perceived naturalness rather than noise suppression. For speaker similarity, we compute the cosine similarity (CosSim, in \%) between embeddings of enhanced and clean speech, extracted using a WavLM-based speaker representation model.\footnote{https://huggingface.co/microsoft/wavlm-base-sv} Intelligibility is measured through the differential word error rate (dWER, in \%), obtained by comparing Wav2Vec2\footnote{https://huggingface.co/facebook/wav2vec2-base-960h} transcriptions of enhanced versus clean speech. 
In addition to non-intrusive metrics used for SE, we use the intrusive measures PESQ \cite{Pesq} and ESTOI \cite{estoi} to assess clean-speech reconstruction after training.
Finally, to quantify the computation needed for inference, we report the number of floating-point operations (FLOPs) required to generate one second of speech, excluding the NAC encoder and decoder. For reference, the NAC encoder and decoder, respectively, require $25$ GFLOPs and $87$ GFLOPs.

\interfootnotelinepenalty=10000

\myparagraph{Baselines}
We compare the proposed models against discriminative approaches for SE. We include several widely adopted discriminative neural networks that learn direct mappings from noisy speech to clean speech in either the time or time-frequency domain: the Deep Complex Convolutional Recurrent Network (DCCRNet) \cite{dccrn}, the Deep Complex U-Net (DCUNet) \cite{dcunet}, the Dual-Path Transformer Network (DPTNet) \cite{dptnet}, and the fully-convolutional Conv-TasNet \cite{luo2019conv}, using models pretrained on Libri1Mix available online.\footnote{https://huggingface.co/JorisCos}
In addition, we consider AnCoGen \cite{sadok2025ancogen}, a model designed for the analysis and controllable generation of speech, in particular SE.\footnote{Contrary to \cite{sadok2025ancogen}, we use the most recent DNSMOS P.835 model for performance evaluation, which explains the difference in reported scores.} To the best of our knowledge, no NAC-based SE methods have publicly released the code or model weights, which prevents direct comparison with such approaches.

\subsection{Results}    

Table~\ref{RES1} reports objective evaluations across all models. The most noticeable observation is that models trained to predict continuous codec representations consistently outperform their discrete counterparts, irrespective of AR/NAR design (on average, +0.80 UTMOS and +0.40 SIG, comparing C-AR, C-NAR, and C-FT with their discrete counterparts). This gap is also evident when comparing \textbf{C-FT} against \textbf{D-FT}.

To further investigate the weakness of discrete models, we introduce a variant of \textbf{D-NAR} that uses continuous inputs $\bar{\mathbf{y}}$ instead of discrete noisy tokens $\mathbf{y}$. This allows us to measure to what extent the input representation affects the \textbf{D-NAR} model. This model (denoted by D-NAR$^\star$ in Table~\ref{RES1}) performs better than other discrete variants, but still lags behind continuous models, suggesting that conditioning on discrete tokens is only part of the problem and that the main bottleneck lies in the output space and associated loss function used for discrete prediction. Additionally, whether continuous or discrete, AR models tend to achieve higher quality (DNSMOS, UTMOS) than NAR models, likely due to their ability to capture temporal dependencies during generation. However, they demonstrate degraded speaker similarity and intelligibility, which could be explained by error accumulation over autoregressive steps. Moreover, the quality gains of continuous AR over continuous NAR are modest compared to the significant computational overhead and the higher dWER values.

We also experiment with fine-tuning the NAC encoder alongside training a \textbf{C-NAR} enhancement model. We chose the C-NAR system to explore this fine-tuning strategy because it already provides a strong balance between quality and efficiency. This strategy, denoted by C-NAR-FT in Table~\ref{RES1}, yields the best compromise between speech quality, intelligibility, and inference speed.

We also observe that fine-tuning the encoder for SE can compromise the primary role of the NAC as a high-fidelity speech codec, since the modified encoder may no longer reconstruct clean speech accurately. To study this effect, we test the \textbf{C-NAR}, \textbf{C-FT}, and \textbf{C-NAR-FT} models on clean speech, and report $\Delta$PESQ and $\Delta$ESTOI relative to the reference NAC before fine-tuning. Results show that C-FT suffers the largest degradation ($\Delta$PESQ $= -0.73$, $\Delta$ESTOI $=-0.03$), C-NAR-FT also degrades reconstruction ($-0.64$, $-0.03$), while C-NAR preserves fidelity best ($-0.32$, $-0.01$). From a practical perspective, this suggests different operating points depending on the target application. For telecommunication scenarios where compression and enhancement must coexist, and clean reconstruction remains important, C-NAR is the preferred choice. For applications where SE performance is the sole priority, the C-NAR-FT model is the preferred choice. % provides good performance while being relatively more efficient. 

%Finally, we include an STFT-based baseline, obtained by replacing $\bar{\mathbf{y}}$ in the C-NAR model with  $STFT(\mathbf{y}_{\text{wav}})$, and training the network to predict a time–frequency mask. 
Finally, we perform a last experiment in which the C-NAR model is trained to predict a time-frequency mask from the STFT of the noisy speech signal. All other model hyperparameters remain the same as before, including the architecture and the MSE loss function. As reported in Table~\ref{RES1}, this \textbf{STFT-NAR} model yields the weakest performance across all metrics (except dWER), confirming the advantage of working in the NAC's latent space.

\section{Conclusion}
In this work, we investigated SE models in the latent space of NACs, comparing discrete and continuous representations, AR and NAR models, and the impact of codec encoder fine-tuning. Our results highlight the advantages of continuous representations and NAR models, as well as the trade-offs between enhancement performance and codec fidelity. 
The study was conducted using only a few hundred hours of training data, offering a complementary perspective to related works that often leverage thousands of hours~\cite{Genhancer,masksr, gense}. However, scaling effects may alter the trends we observed in this study. Also, we did not explore the use of semantic tokens, which are also frequently employed in the previously mentioned literature and could yield different conclusions. Finally, our analysis is restricted to the Descript Audio Codec, extending the study to other NACs would help assess the generality of our findings.\\

\myparagraph{Acknowledgments} This work was performed using computational resources from the Mésocentre computing center\footnote{\texttt{https://mesocentre.universite-paris-saclay.fr}} of Université Paris-Saclay, CentraleSupélec, and  Ecole Normale Supérieure Paris-Saclay, as part of the DEGREASE project (ANR-23-CE23-0009), funded by the French National Research Agency.

\bibliographystyle{IEEEbib}

\begingroup
\fontsize{8pt}{10.5pt}\selectfont 
\bibliography{strings_initials}
\endgroup

\end{document}